\documentclass[pre,twocolumn,amsmath,amssymb,showpacs]{revtex4}

\usepackage{graphicx}
\usepackage{latexsym}
\usepackage{amsmath}
\usepackage{amssymb}
\usepackage{amsfonts}
\usepackage{bm}

\newcommand{\la}{\left<}
\newcommand{\ra}{\right>}

\newcommand{\ddiff}{\ensuremath{\text{d}}}
\newcommand{\Trace}{\ensuremath{\text{Tr}}}
\newcommand{\sij}{\ensuremath{s_{l}}}

\newcommand{\sija}{\ensuremath{s_{l}^{\alpha}}}
\newcommand{\sijb}{\ensuremath{s_{l}^{\beta}}}

\newcommand{\nija}{\ensuremath{n_{l}^{\alpha}}}
\newcommand{\nijb}{\ensuremath{n_{l}^{\beta}}}
\newcommand{\nijc}{\ensuremath{n_{l}^{\gamma}}}
\newcommand{\nijd}{\ensuremath{n_{l}^{\delta}}}

\newcommand{\kB}{\mbox{$k_{\rm B}$}}
\newcommand{\kBT}{\mbox{$k_{\rm B}T$}}

\newcommand{\uprime}{\ensuremath{u^{\prime}}}
\newcommand{\uprimeprime}{\ensuremath{u^{\prime\prime}}}
\newcommand{\utrunc}{\ensuremath{u_\mathrm{t}}}
\newcommand{\utruncprime}{\ensuremath{u^{\prime}_\mathrm{t}}}

\newcommand{\ushift}{\ensuremath{u_\mathrm{s}}}
\newcommand{\ushiftprime}{\ensuremath{u^{\prime}_\mathrm{s}}}
\newcommand{\ushiftprimeprime}{\ensuremath{u^{\prime\prime}_\mathrm{s}}}

\newcommand{\Pid}{\ensuremath{P_\mathrm{id}}}
\newcommand{\Pex}{\ensuremath{P_\mathrm{ex}}}
\newcommand{\Pexuncorr}{\ensuremath{\tilde{P}_\mathrm{ex}}}
\newcommand{\Pexcut}{\ensuremath{\Delta P_\mathrm{ex}}}
\newcommand{\Kcut}{\ensuremath{\Delta K}}
\newcommand{\Gcut}{\ensuremath{\Delta G}}
\newcommand{\Kuncorr}{\ensuremath{\tilde{K}}}
\newcommand{\Guncorr}{\ensuremath{\tilde{G}}}
\newcommand{\muB}{\ensuremath{\mu_\mathrm{B}}}
\newcommand{\muBcut}{\ensuremath{\Delta \mu_\mathrm{B}}}
\newcommand{\muF}{\ensuremath{\mu_\mathrm{F}}}
\newcommand{\laB}{\ensuremath{\lambda_\mathrm{B}}}
\newcommand{\laBcut}{\ensuremath{\Delta \lambda_\mathrm{B}}}
\newcommand{\laF}{\ensuremath{\lambda_\mathrm{F}}}
\newcommand{\CKabcd}{\ensuremath{C_\mathrm{K}^{\alpha\beta\gamma\delta}}}
\newcommand{\CBabcd}{\ensuremath{C_\mathrm{B}^{\alpha\beta\gamma\delta}}}
\newcommand{\CBabcduncorr}{\ensuremath{\tilde{C}_\mathrm{B}^{\alpha\beta\gamma\delta}}}
\newcommand{\CFabcd}{\ensuremath{C_\mathrm{F}^{\alpha\beta\gamma\delta}}}

\newcommand{\pressTen}{\ensuremath{\hat{P}_\mathrm{ex}^{\alpha\beta}}}

\newcommand{\histoP}{\ensuremath{h_\mathrm{1}}}
\newcommand{\histoB}{\ensuremath{h_\mathrm{2}}}
\newcommand{\histoBabcd}{\ensuremath{h_\mathrm{2}^{\alpha\beta\gamma\delta}}}

\newcommand{\rcut}{\ensuremath{r_\mathrm{c}}}
\newcommand{\scut}{\ensuremath{s_\mathrm{c}}}
\newcommand{\scutm}{\ensuremath{s_\mathrm{c}^{-}}}
\newcommand{\smin}{\ensuremath{s_\mathrm{0}}}

\newcommand{\uLJ}{\ensuremath{u_\mathrm{LJ}}}
\newcommand{\fLJ}{\ensuremath{f_\mathrm{LJ}}}
\newcommand{\nA}{\ensuremath{n_\mathrm{A}}}
\newcommand{\nB}{\ensuremath{n_\mathrm{B}}}

\newcommand{\epsAA}{\ensuremath{\epsilon_\mathrm{AA}}}
\newcommand{\epsBB}{\ensuremath{\epsilon_\mathrm{BB}}}
\newcommand{\epsAB}{\ensuremath{\epsilon_\mathrm{AB}}}
\newcommand{\epsij}{\ensuremath{\epsilon_{l}}}
\newcommand{\sigAA}{\ensuremath{\sigma_\mathrm{AA}}}
\newcommand{\sigBB}{\ensuremath{\sigma_\mathrm{BB}}}
\newcommand{\sigAB}{\ensuremath{\sigma_\mathrm{AB}}}
\newcommand{\sigmal}{\ensuremath{\sigma_{l}}}

\newcommand{\sigmaij}{\ensuremath{\sigma_{l}}}

\newcommand{\ca}{\ensuremath{c_{a}}}
\newcommand{\cb}{\ensuremath{c_{b}}}
\newcommand{\diffDi}{\ensuremath{\ddiff t}}
\newcommand{\diffDj}{\ensuremath{\ddiff t^{\prime}}}
\newcommand{\pDi}{\ensuremath{c_{t}}}
\newcommand{\pDj}{\ensuremath{c_{t^{\prime}}}}
\newcommand{\wDiDj}{\ensuremath{w_{tt^{\prime}}}}
\newcommand{\vDiDj}{\ensuremath{v_{tt^{\prime}}}}
\newcommand{\gDiDj}{\ensuremath{g_{tt^{\prime}}}}
\newcommand{\sigref}{\ensuremath{\sigma_{\mathrm{ref}}}}
\newcommand{\epsref}{\ensuremath{\epsilon_{\mathrm{ref}}}}

\newcommand{\msd}{\ensuremath{h}}

\begin{document}

\title{Impulsive correction to the elastic moduli obtained using the\\ 
stress-fluctuation formalism in systems with truncated pair potential}

\author{H.~Xu}
\affiliation{LCP-A2MC, Institut Jean Barriol, Universit\'e de Lorraine \& CNRS, 1 bd Arago, 57078 Metz Cedex 03, France}
\author{J.P.~Wittmer}
\email{joachim.wittmer@ics-cnrs.unistra.fr}
\affiliation{Institut Charles Sadron, Universit\'e de Strasbourg \& CNRS, 23 rue du Loess, 67034 Strasbourg Cedex, France}
\author{P.~Poli\'nska}
\affiliation{Institut Charles Sadron, Universit\'e de Strasbourg \& CNRS, 23 rue du Loess, 67034 Strasbourg Cedex, France}
\author{J. Baschnagel}
\affiliation{Institut Charles Sadron, Universit\'e de Strasbourg \& CNRS, 23 rue du Loess, 67034 Strasbourg Cedex, France}

\begin{abstract}
The truncation of a pair potential at a distance $\rcut$ is well-known to imply in general an impulsive 
correction to the pressure and other moments of the first derivatives of the potential. 
That depending on $\rcut$ the truncation may also be of relevance to higher derivatives is shown 
theoretically for the Born contributions to the elastic moduli obtained using the stress-fluctuation 
formalism in $d$ dimensions. 
Focusing on isotropic liquids for which the shear modulus $G$ must vanish by construction, 
the predicted corrections are tested numerically for binary mixtures and polydisperse Lennard-Jones beads
in, respectively, $d=3$ and $d=2$ dimensions.
\end{abstract}

\date{\today}
\maketitle

\section{Introduction}
\label{sec_intro}

\paragraph*{Background.}
It is common practice in computational condensed matter physics 
\cite{AllenTildesleyBook,FrenkelSmitBook,ThijssenBook} 
to truncate a pair interaction potential $U(r)$ at a conveniently chosen cutoff distance $\rcut$ 
with $r$ being the distance between two particles $i$ and $j$.
This allows to reduce the number of interactions to be computed
--- energy or force calculations become thus ${\cal O}(n)$-processes with $n$ being the particle number ---
but introduces some technical difficulties, e.g., instabilities in the numerical solution of differential equations 
as well-studied in the past especially for the molecular dynamics (MD) method \cite{AllenTildesleyBook,ToxvaerdDyre11}.
Let us label the interaction between two beads $i < j$ by an index $l$.
For simplicity of the presentation and without restricting much in practice
the generality of our results, it is assumed below that
\begin{itemize}
\item
the pair potential scales as $U(r) \equiv u(s)$ with the reduced dimensionless distance 
$s=r/\sigmaij$ where $\sigmaij$ characterizes the range of the interaction $l$ and
\item
the same reduced cutoff $\scut = \rcut/\sigmaij$ is set for all interactions $l$. 
\end{itemize}
For monodisperse beads with constant bead diameter $\sigma$, 
as for the standard Lennard-Jones (LJ) potential \cite{AllenTildesleyBook}
\begin{equation}
\uLJ(s) = 4 \epsilon \left(\frac{1}{s^{12}} - \frac{1}{s^6} \right),
\label{eq_LJ}
\end{equation}
the scaling variable becomes simply $s=r/\sigma$ and the reduced cutoff $\scut = \rcut/\sigma$.
Even if the truncated potential 
\begin{equation}
\utrunc(s) = u(s) H(\scut -s)
\label{eq_utrunc}
\end{equation}
with $H(s)$ being the Heaviside function \cite{abramowitz} is taken {\em by definition} as the new system Hamiltonian, 
it is well known that impulsive corrections at the cutoff have to be taken into account in general
for the pressure $P$ and other moments of the first derivatives of the potential
\cite{FrenkelSmitBook}.
These corrections can be avoided of course by considering a properly shifted potential \cite{FrenkelSmitBook}
\begin{equation}
\ushift(s) = \left( u(s) - u(\scut) \right) H(\scut -s)
\label{eq_ushift}
\end{equation}
as emphasized below (Sec.~\ref{theo_shifted}).

\paragraph*{Goal of presented work.}
In this report we wish to remind that the standard shifting of a truncated potential is insufficient 
in general to avoid impulsive corrections for moments of second (and higher) derivatives of the potential. 
We demonstrate here that this is particulary the case for the Born contribution $\CBabcd$ (defined below) 
to the elastic moduli computed using the stress-fluctuation formalism described in great detail in the literature
\cite{RowlinsonBook,FrenkelSmitBook,Hoover69,Ray84,Ray88,Lutsko88,Lutsko89,Lemaitre04,Pablo04,Pablo05,Barrat06,SBM11,papthermoflex}.
This should be of importance for the precise localization of the transition between
different thermodynamic phases using the elastic moduli, especially for liquid ($G=0$) to solid ($G > 0$) 
transitions in network forming systems where the shear modulus $G$ plays the role of an order parameter 
\cite{Zippelius06}. This is the case, e.g. for colloidal gels \cite{Kob08}, 
hyperbranched polymer chains with sticky end-groups \cite{Friedrich10} or 
bridged networks of telechelic polymers in water-oil emulsions \cite{Porte01b,Porte03} or
living polymer-like micellar systems \cite{Ligoure07}.

\paragraph*{Outline.}
The paper is organized as follows: After reminding first in Sec.~\ref{theo_reminder} the known corrections 
for the pressure and similar first derivates of the potential, the impulsive correction for the general 
Born contribution $\CBabcd$ is stated in Sec.~\ref{theo_key}. 
We describe then in Sec.~\ref{theo_isotropic} the corrections 
for the compression modulus $K$ and the shear modulus $G$ in isotropic systems. 
We comment on polydispersity effects and mixed potentials in Sec.~\ref{theo_mixtures}. 
Our results are rephrased in terms of the radial pair distribution function $g(s)$ in Sec.~\ref{theo_gr}
which allows to predict the asymptotic behavior for large $\scut$.
Section~\ref{sec_algo} gives some technical details on the two numerical model systems \cite{Kob95,TWLB02}
in $d=3$ and $d=2$ dimensions used to test our predictions in Sec.~\ref{sec_res}.
We consider in this paper the liquid high-temperature regime of both (in principle glass-forming) 
models where the shear modulus $G$ must vanish \cite{RowlinsonBook,HansenBook}, 
since this provides a simple reference for testing the predicted corrections.

\section{Theoretical predictions}
\label{sec_theo}

\subsection{Reminder}
\label{theo_reminder}

\paragraph*{Truncated potential.}
As usual for pairwise additive interactions the mean pressure $P = \Pid + \Pex$ may
be obtained as the sum of the ideal kinetic contribution $\Pid = \kBT \rho$
and the excess pressure contribution \cite{HansenBook}
\begin{equation}
\Pex = - \frac{1}{d V} \sum_{l} \la \sij \utruncprime(\sij) \ra 
\label{eq_Pex}
\end{equation}
with $\rho=n/V$ being the number density, $n$ the particle number,
$V$ the $d$-dimensional volume and $\la \ldots \ra$ indicating the usual 
thermal average over the configuration ensemble.
(A prime denotes a derivative of a function with respect to its argument.)
By taking the derivative of the truncated potential 
\begin{equation}
\utruncprime(s) = \uprime(s) H(\scut -s) - u(s) \delta(s-\scut)
\label{eq_utruncprime}
\end{equation}
the excess pressure may be written as the sum $\Pex = \Pexuncorr + \Pexcut$ of an uncorrected (bare) 
contribution $\Pexuncorr$ and an impulsive correction $\Pexcut$. The latter correction is obtained 
numerically from \cite{FrenkelSmitBook}
\begin{eqnarray}
\Pexcut & = & \lim_{s\to \scutm} \histoP(s) \mbox{ with } \nonumber\\
\histoP(s) & \equiv & \frac{1}{d V} \sum_{l} \la \sij u(\sij) \ \delta(\sij-s) \ra
\label{eq_histoP}
\end{eqnarray}
being a weighted histogram. 
In practice, the proper limit $s \to \scutm$ may be replaced by setting $s = \scut$ for the histogram.

\paragraph*{Shifted potential.}
\label{theo_shifted}
The impulsive correction related to first derivatives of the truncated potential can be avoided  
by considering the shifted potential $\ushift(s)$, Eq.~(\ref{eq_ushift}), since
$\ushiftprime(s) = \uprime(s) H(\scut -s)$.
With this choice {\em no} impulsive correction arises either for similar observables such
as, e.g., moments of the instantaneous excess pressure tensor 
\begin{eqnarray}
\pressTen & = & -\frac{1}{V} \sum_{l} \sija \frac{\partial \ushift(\sij)}{\partial \sijb} \nonumber \\
& = & -\frac{1}{V} \sum_{l} \sij \ushiftprime(\sij) \ \nija \nijb.
\label{eq_pressTen}
\end{eqnarray}
Here $\sija$, $\ldots$ stand for the spatial components of the reduced distance between the particles,
$\nija = \sija/\sij$, $\ldots$ for the corresponding components of the normalized distance vector and
Greek letters are used for the spatial coordinates $\alpha,\beta,\gamma,\delta = 1,\ldots,d$.
(We remind that the mean excess pressure $\Pex$ is the averaged trace over the 
instantaneous excess pressure tensor, $\Pex = < \Trace[\pressTen] >/d$.)
Specifically, if the potential is shifted, all impulsive corrections are avoided for the 
excess pressure fluctuations 
\begin{equation}
\CFabcd \equiv - \beta V \left( \la \hat{P}_\mathrm{ex}^{\alpha\beta} \hat{P}_\mathrm{ex}^{\gamma\delta} \ra - 
\la \hat{P}_\mathrm{ex}^{\alpha\beta} \ra \la \hat{P}_\mathrm{ex}^{\gamma\delta} \ra \right)
\label{eq_muF}
\end{equation}
($\beta \equiv 1/\kB T$ being the inverse temperature)
which give important contributions 
--- especially for polymer-type liquids \cite{SBM11,papthermoflex}
and amorphous solids \cite{Lemaitre04,Barrat06} ---
to the elastic moduli computed using the stress-fluctuation formalism 
\cite{FrenkelSmitBook}. 

\subsection{Key point made}
\label{theo_key}

\paragraph*{Correction to the Born term.}
An even more important contribution to the elastic moduli (especially at high densities) is given
by the Born term $\CBabcd$ already mentioned in the Introduction \cite{foot_CKabcd}. Being a moment 
of the first and the second derivatives of the potential it is defined as 
\cite{FrenkelSmitBook,Lutsko89,SBM11,TWLB02}
\begin{equation}
\CBabcd = \frac{1}{V} \sum_{l} \la \left( \sij^2 \ushiftprimeprime(\sij) - \sij \ushiftprime(\sij) \right)
\nija \nijb \nijc \nijd \ra
\label{eq_CBabcd}
\end{equation}
using the notations given above.
We remind that for solids with well-defined reference positions and displacement fields the Born 
contribution is known to describe the (free) energy change assuming an {\em affine} response to an 
imposed homogeneous strain \cite{Lutsko89,Lemaitre04,Barrat06,SBM11,TWLB02}. 
Assuming now a truncated and shifted potential the impulsive correction $\Delta \CBabcd$
to $\CBabcd = \CBabcduncorr + \Delta \CBabcd$ is simply obtained using
\begin{equation}
\ushiftprimeprime(s) =  \uprimeprime(s) H(\scut-s) - \uprime(s) \delta(s-\scut)
\label{eq_ushiftprimeprime}
\end{equation}
which yields
\begin{eqnarray}
\Delta \CBabcd & = & - \lim_{s\to \scutm} \histoBabcd(s) \mbox{ with } 
\label{eq_CBabd_correct} \\
\histoBabcd(s) & \equiv & \frac{1}{V} \sum_{l} \la  \sij^2 \uprime(\sij) 
\nija \nijb \nijc \nijd \ \delta(\sij-s) \ra. 
\nonumber
\end{eqnarray}

\paragraph*{General impulsive correction.}
More generally, one might consider a property
$$A = \frac{1}{V} \sum_l \la f(\sij) \ushift^{(n)}(\sij) \ra$$
with $f(s)$ being a specified function and $(n)$ denoting
the $n$-th derivative of the shifted potential $\ushift(s)$.
Let us further suppose that all potential derivatives up to the
$(n-2)$-th one do vanish at the cutoff $\scut$.
It thus follows that $A = \tilde{A} + \Delta A$ takes
an impulsive correction
\begin{eqnarray}
\Delta A & = & -\lim_{s \to \scutm} h_n(s) \mbox{ with } \nonumber \\
h_n(s) & \equiv & \frac{1}{V} \sum_l \la f(\sij) u^{(n-1)}(\sij) \ \delta(\sij -s) \ra
\label{eq_hn_histo}
\end{eqnarray}
being the relevant histogram. 

\paragraph*{Generalized shifting.}
Obviously, the original potential may not only be shifted by a constant $u(\scut)$ but by a polynomial of $s$ 
to make vanish the first and arbitrarily high derivatives of the potential at $s=\scut$. In this way all 
impulsive corrections could be avoided in principle.
Since discontinuous forces at the cutoff may cause problems in MD simulations, a number of studies use for 
instance a ``shifted-force potential" where a linear term is added to the potential 
\cite{AllenTildesleyBook,ToxvaerdDyre11}.
The difference between the original potential and the generalized shifted potential removing the cutoff 
discontinuities means, of course, that the computed properties deviate to some extend from the original model.
Only if the generalized shifting is weak, one may recover the correct thermodynamic properties 
using a first-order perturbation scheme \cite{AllenTildesleyBook}.
Since the (simply) shifted potential $\ushift(s)$, Eq.~(\ref{eq_ushift}), is anyway the most common choice
\cite{Hoover69,Ray84,Ray88,Lemaitre04,Pablo04,Pablo05,Barrat06,SBM11,papthermoflex,Kob95,TWLB02}, 
we restrict the presentation on this case and demonstrate how the impulsive correction associated 
to the non-vanishing $\ushiftprime(\scutm)$ can readily be computed.

\subsection{Isotropic systems}
\label{theo_isotropic}

\paragraph*{Lam\'e coefficients.}
In order to show that these corrections may be of relevance 
we focus now on homogeneous and isotropic systems. We remind first that the two elastic Lam\'e coefficients 
$\lambda$ and $\mu$ characterizing the elastic properties of such systems may be computed numerically using
\cite{SBM11,papthermoflex}
\begin{eqnarray}
\lambda & = & \laF  + \laB, \nonumber \\
\mu - \Pid  & = & \muF + \muB \label{eq_lamu2BF}
\end{eqnarray}
where the only contribution due to the kinetic energy of the particles is contained 
by the ideal gas pressure $\Pid$ indicated for $\mu$ \cite{foot_kinelast}. 
The first contributions indicated on the right hand-side of Eq.~(\ref{eq_lamu2BF}) are the excess pressure 
fluctuation contributions $\laF$ and $\muF$ which may be obtained from the general $\CFabcd$ by setting, e.g.,
$\alpha=\beta=1$ and $\gamma=\delta=2$ for $\laF$ and
$\alpha=\gamma=1$ and $\beta=\delta=2$ for $\muF$ characterizing the shear stress fluctuations. 
The so-called ``Born-Lam\'e coefficients" \cite{SBM11}
\begin{equation}
\laB \equiv \muB  
\equiv \frac{1}{d(d+2) V} \sum_{l} \la \left( \sij^2 u^{\prime\prime}(\sij) - \sij u^{\prime}(\sij) \right) \ra  
\label{eq_lameBorn}
\end{equation}
may be obtained from the general Born terms $\CBabcd$ by setting, 
e.g., $\alpha=\gamma=1$ and $\beta=\delta=2$. 
The $d$-dependent prefactor stems from the assumed isotropy of the system
and the mathematical formula 
\begin{equation}
\la \left( \nija \nijb \right)^2 \ra =
\frac{1}{d(d+2)} \left(1 + 2 \delta_{\alpha\beta} \right)
\label{eq_dprefactor}
\end{equation}
($\delta_{\alpha\beta}$ being the Kronecker symbol \cite{abramowitz})
for the components of a unit vector in $d$ dimensions pointing into arbitrary directions.
Equation~(\ref{eq_CBabd_correct}) implies then an impulsive correction
\begin{eqnarray}
\laBcut & = & \muBcut = - \lim_{s\to \scutm} \histoB(s) \mbox{ with } \nonumber\\
\histoB(s) & \equiv & \frac{1}{d (d+2) V} \sum_{l} \la  \sij^2 \uprime(\sij) \ \delta(\sij-s) \ra.
\label{eq_histoLame}
\end{eqnarray}

\paragraph*{Compression and shear modulus.}
Instead of using the Lam\'e coefficients it is from the experimental point of view
more natural to characterize isotropic bodies using the compression modulus $K$
and the shear modulus $G$. The latter moduli may be expressed as
\begin{eqnarray}
K & = & (\lambda + P) + \frac{2}{d} G, \label{eq_lame2K}\\
G & = & \mu - P = \muB + \muF - \Pex. \label{eq_lame2G} 
\end{eqnarray}
We follow here the notation of Ref.~\cite{SBM11} to emphasize the explicit pressure dependence 
which is often (incorrectly) omitted as clearly pointed out by Birch \cite{Birch37} and Wallace \cite{Wallace70}.
As one expects, kinetic elastic contributions terms do not enter explicitly for the shear modulus.
Since only the Born contributions $\laB=\muB$ cause a cutoff correction,
this implies 
$K = \Kuncorr + \Kcut$ and $G = \Guncorr + \Gcut$ with 
$\Kuncorr$ and $\Guncorr$ being the uncorrected (bare) moduli and
\begin{eqnarray}
\Kcut & = & \laBcut + \frac{2}{d} \muBcut =  \frac{2+d}{d} \muBcut, \label{eq_deltaK} \\
\Gcut & = & \muBcut \label{eq_deltaG} 
\end{eqnarray}
the impulsive corrections. We shall test these predictions numerically in Sec.~\ref{sec_res}.
%

\subsection{Polydispersity and mixed potentials}
\label{theo_mixtures}

As stated in the Introduction we assume throughout this work the scaling $U(r) \equiv u(s)$ of the 
pair potential in terms of the reduced distance $s = r/\sigmaij$. This is done not only for dimensional 
reasons but, more importantly, to describe a broad range of model systems for mixtures and polydisperse 
systems where the interaction range $\sigmaij$ may differ for each interaction $l$. Moreover,
the type and/or the parameter set of the pair potential may vary for different interactions.
For such mixed potentials $u(s)$, $\utrunc(s)$ and $\ushift(s)$ and their derivatives take in principal 
an explicit index $l$, i.e. one should write $u_l(s)$, $u_{\mathrm{t},l}(s)$, $u_{\mathrm{s},l}(s)$ and so on.
This is only not done here to keep a concise notation. For example one might wish to consider
\begin{itemize}
\item
a generic polymer bead-spring model where some interactions $l$ describe the bonded interactions 
between monomers along the chain (which are normally not truncated and need not to be corrected)
and some the excluded volume interactions between the beads \cite{papthermoflex}. 
\item
the generalization of the monodisperse LJ potential, Eq.~(\ref{eq_LJ}),
\begin{equation}
u_l(s) = 4\epsij \left(s^{-12} - s^{-6} \right) \mbox{ with } s = r/\sigmaij
\label{eq_LJgeneralized}
\end{equation}
where $\epsij$ and $\sigmaij$ are fixed for each interaction $l$.
In practise, each particle $i$ may be characterized by an energy scale $E_i$ and a ``diameter" $D_i$. 
The interaction parameters $\epsij(E_i,E_j)$ and $\sigmaij(D_i,D_j)$ are then given in terms of
specified functions of these properties. 
\item
the famous Kob-Andersen (KA) model for binary mixtures of beads of type $A$ and $B$ \cite{Kob95},
a particular case of Eq.~(\ref{eq_LJgeneralized}) with fixed 
interaction ranges $\sigAA$, $\sigBB$ and $\sigAB$ and 
energy parameters $\epsAA$, $\epsBB$ and $\epsAB$ characterizing, respectively,
$AA$-, $BB$- and $AB$-contacts.
\item
a network forming emulsion of oil droplets in water bridged by telechelic polymers 
where the oil droplets are modeled as big LJ spheres, the telechelic polymers by a bead-spring
model with a soluble ``spacer" in the middle of the chain and insoluble end-groups (``stickers") 
strongly attracted by the oil droplets \cite{Porte01b,Porte03}. 
Assuming sufficiently strong (in strength, number and life-time) sticker-oil interactions,
such a system should behave as a soft solid with a finite shear modulus $G$ 
(at least for a fixed finite sampling time) which may be probed, at least in principle, 
using Eq.~(\ref{eq_lame2G}).
\end{itemize}
The impulsive corrections given in Eq.~(\ref{eq_histoP}) for the pressure $P$, 
in Eq.~(\ref{eq_CBabd_correct}) for the general Born term $\CBabcd$ and 
in Eq.~(\ref{eq_histoLame}) for the Born Lam\'e coefficients $\laB=\muB$ in isotropic systems 
stated all in terms of, respectively, the histograms $\histoP(s)$, $\histoBabcd(s)$ and $\histoB(s)$ 
remain indeed valid for such explicitly $l$-dependent potentials. From the numerical point of view, 
this is all what is needed and the direct computation of these histograms remains in all cases 
straightforward as illustrated in Sec.~\ref{res_histo}.

\subsection{Radial pair distribution function $g(r)$}
\label{theo_gr}

\paragraph*{Notations.}
Especially for simple and complex fluids and for all sorts of glass-forming systems 
\cite{RowlinsonBook,HansenBook} 
it is common practice to reexpress correlations and histograms in terms of the 
radial pair distribution function $g(r)$ \cite{AllenTildesleyBook,FrenkelSmitBook}.
This is also of interest here since for large cutoff distances the pair distribution function 
must drop out, $g(\rcut) \to 1$, allowing thus to predict the corrections in this limit.
Let us remind first that, using the Gamma function $\Gamma(x)$ \cite{abramowitz},
the $(d-1)$-dimensional surface of a $d$-sphere of radius $r$ is given by 
\begin{equation}
A(r) = \frac{2 \pi^{d/2}}{\Gamma(d/2)} \ r^{d-1}
\mbox{ for } d = 2,3,\ldots
\label{eq_Adef}
\end{equation}
and similarly for the (dimensionless) surface $A(s)$ using the reduced distance $s$.

\paragraph*{Monodisperse interactions.}
For strictly monodisperse beads and similar interactions of constant interaction range $\sigma$ 
it is seen that Eq.~(\ref{eq_histoP}) for the pressure correction becomes
\begin{equation}
\Pexcut = 
\frac{1}{2} \ \frac{1}{d}
\rho^2 \sigma^d A(\scut) \scut   u(\scut) \times g(\scut)
\label{eq_Pexcutmono}
\end{equation}
where the factor $1/2$  assures that every interaction is only counted once.
Note that we have set $g(s) \equiv g(r/\sigma)$ without introducing a new symbol.
For the LJ potential, Eq.~(\ref{eq_LJ}), this leads to
\begin{equation}
\Pexcut = - \frac{4 \pi^{d/2}}{\Gamma(d/2)d} \rho^2 \sigma^d \epsilon \scut^{d-6} (1-\scut^{-6}) \times g(\scut).
\label{eq_PexcutmonoLJ}
\end{equation}
Please note that the expression given in Ref.~\cite{FrenkelSmitBook} is recovered by setting $d=3$ and
assuming $g(\scut) \approx 1$.
Similarly, one obtains from Eq.~(\ref{eq_histoLame}) the correction
\begin{equation}
\muBcut =  - 
\frac{1}{2} \ \frac{1}{d (d+2)}
\rho^2 \sigma^{d} A(\scut) \scut^{2} \uprime(\scut) \times g(\scut)
\label{eq_muBcutmono}
\end{equation}
for the Born-Lam\'e coefficient we are mainly interested in. 
For a LJ potential this becomes
\begin{equation}
\muBcut = - \frac{24\pi^{d/2}}{d(d+2)\Gamma(d/2)} 
\rho^2 \sigma^d \epsilon \fLJ(\scut) \times g(\scut)
\label{eq_muBcutmonoLJ}
\end{equation}
where we have defined 
\begin{equation}
\fLJ(s) \equiv (1 - (\smin/s)^6)/s^{6-d}
\label{eq_fLJdef}
\end{equation}
with $\smin = 2^{1/6}$ being the minimum of the potential.
For sufficiently large cutoff distances where $g(\scut) \approx 1$
the correction thus decays as 
\begin{equation}
\muBcut \sim - A(\scut) \scut^{2} \uprime(\scut),
\label{eq_muBcut_asymp}
\end{equation}
e.g. $\muBcut \sim - 1/\scut^{6-d}$ for a LJ potential.
This asymptotic behavior also holds for the more complicated cases discussed below. 

\paragraph*{Mixtures.}
Many experimental relevant systems have mixed potentials such as the KA model for binary 
colloidal mixtures sketched above. In general the interaction potential $U_{ab}(r)=u_{ab}(s)$ between 
beads of two species $a$ and $b$ takes different energy parameters which causes different weights
at the cutoff depending on which particles interact.
The impulsive corrections of such mixtures are readily obtained by linear superposition of 
Eq.~(\ref{eq_muBcutmono}) for different contributions $(a,b)$. 
Let $\ca = \rho_a/\rho$ denote the mole fraction of species $a$,
$\sigma_{ab}$ the interaction range between a bead of type $a$ and a bead of type $b$ and
$g_{ab}(s)$ the respective radial pair distribution function. 
The impulsive correction to the Born-Lam\'e coefficient thus becomes
\begin{eqnarray}
\muBcut & = & - \frac{1}{2} \frac{1}{d (d+1)} \rho^2 A(\scut) \scut^{2} \nonumber \\
 & \times &
\sum_{a} \sum_{b} \ca \cb \sigma_{ab}^d \uprime_{ab}(\scut) \ g_{ab}(\scut)
\label{eq_muBcut_mixture}
\end{eqnarray} 
where we have used that for all types of interaction we have the {\em same} reduced cutoff $\scut$.

Let us now assume a mixture described by the generalized LJ potential 
$u_{ab}(s) = \epsilon_{ab} (1/s^{12}-1/s^6)$ with $s=r/\sigma_{ab}$.
A reference energy $\epsref$ and a reference interaction range $\sigref$
may arbitrarily be defined using, say, the interaction of two beads of type $a=b=1$,
i.e. $\epsref \equiv \epsilon_{11}$ and $\sigref \equiv \sigma_{11}$.
Defining the dimensionless ratios $w_{ab} \equiv \epsilon_{ab}/\epsref$ and $v_{ab} = (\sigma_{ab}/\sigref)^d$ 
we may thus rewrite the general Eq.~(\ref{eq_muBcut_mixture}) as
\begin{eqnarray}
\muBcut & = & - \frac{24 \pi^{d/2}}{d(d+2)\Gamma(d/2)} \rho^2 \sigref^d \epsref \ \fLJ(\scut) \nonumber \\
 & \times & \sum_a \sum_b \ca \cb v_{ab} w_{ab} \ g_{ab}(\scut).
\label{eq_muBcut_mixtureLJ}
\end{eqnarray}
Since $g_{ab}(\scut) \to 1$ for large $\scut$, the function $\fLJ(\scut)$ determines the scaling 
as already stated above, Eq.~(\ref{eq_muBcut_asymp}).

\paragraph*{Continuous polydispersity.}
We turn now to systems with a continuous polydispersity as in the second model investigated numerically below.
Let us assume that each bead is characerized by a bead diameter $D$ which is distributed according to a 
well-defined normalized distribution $\pDi$ with $t = D/\sigref$ being a reduced bead diameter with 
respect to some reference length $\sigref$. 
To be specific we shall assume a generalized LJ potential, Eq.~(\ref{eq_LJgeneralized}),
where the interaction range $\sigma_{tt^{\prime}}$ and the energy scale $\epsilon_{tt^{\prime}}$ 
between two beads are uniquely specified by the two reduced diameters $t$ and $t^{\prime}$. 
Defining  $\wDiDj = \epsilon_{tt^{\prime}}/\epsref$,
$\vDiDj = \left( \sigma_{tt^{\prime}}/\sigref \right)^d$ and
using the radial pair distribution function $\gDiDj(s)$ for two beads of reduced diameter $t$ and $t^{\prime}$,
the double-sum in Eq.~(\ref{eq_muBcut_mixtureLJ}) can be rewritten as the double-integral
\begin{eqnarray}
\muBcut & = & - \frac{24 \pi^{d/2}}{d(d+2)\Gamma(d/2)} \rho^2 \sigref^d \epsref \ \fLJ(\scut) \nonumber\\
 & \times & \int \diffDi \int \diffDj \ \pDi \pDj \vDiDj \wDiDj \ \gDiDj(\scut).
\label{eq_muBcut_polyLJ}
\end{eqnarray}
In order to determine $\muBcut$ from Eq.~(\ref{eq_muBcut_polyLJ}) one needs to prescribe
the laws for $\pDi$,  $\sigma_{tt^{\prime}}$ and $\epsilon_{tt^{\prime}}$. In the large-$\scut$
limit the double-integral becomes in any case constant, i.e. we have again
$\muBcut \sim - \fLJ(\scut) \sim - 1/\scut^{6-d}$. 
%

\begin{table}[t]
\begin{tabular}{|c||c|c|c|c|c|c||c|}
\hline
$\scut/\smin$         &
$e \beta$             & 
$P \beta/\rho$        & 
$\Kuncorr \beta/\rho$ &
$K \beta/\rho$        &
$\Guncorr \beta/\rho$ &
$\muBcut \beta/\rho$  &   
$K \beta/\rho$        
\\ \hline
0.9 & 0.162 & 4.61 & -19.2& 15.0 &-17.06& 17.08 & 14.5\\
1.0 & 0.329 & 5.39 & 19.0 & 19.0 & 0.05 &  0.03 & 18.8 \\
1.1 & 0.103 & 4.90 & 20.9 & 17.2 & 1.86 & -1.86 & 17.2 \\
1.5 & -1.24 & 3.13 & 14.1 & 13.2 & 0.44 & -0.43 & 13.3 \\
2.0 & -1.69 & 2.71 & 13.0 & 12.4 & 0.28 & -0.30 & 12.3 \\
2.5 & -1.83 & 2.55 & 12.3 & 12.1 & 0.09 & -0.10 & 12.0\\
3.0 & -1.89 & 2.50 & 12.1 & 12.0 & 0.05 & -0.05 & 11.9\\
3.5 & -1.91 & 2.47 & 12.1 & 12.0 & 0.03 & -0.03 & 12.1\\
4.0 & -1.92 & 2.46 & 11.9 & 11.8 & 0.03 & -0.02 & 11.9\\
\hline
\end{tabular}
\vspace*{0.5cm}
\caption[]{Various properties for polydisperse LJ beads at temperature $T=1$ and density $\rho \approx 0.72$
{\em vs.} the reduced cutoff distance $\scut/\smin$ with $\smin=2^{1/6}$ being the minimum of the potential:
energy per bead $e$,
total pressure $P = \Pid + \Pex$,
uncorrected compression modulus $\Kuncorr$,
corrected compression modulus $K = \Kuncorr + 2 \muBcut$,
bare shear modulus $\Guncorr$ and
impulsive correction $\muBcut$ obtained from the histogram $\histoB(s)$ at $s = \scut$.
The corrected shear modulus $G = \Guncorr + \Gcut$ vanishes as it should.
The last column refers to the compression modulus $K$ obtained using Eq.~(\ref{eq_Kisobaric}) 
for isobaric ensembles kept at the same pressure $P$ (third column).
\label{tab_scut}}
\end{table}

\section{Computational issues}
\label{sec_algo}
To illustrate the above predictions we present computational data using two extremely well studied models
of colloidal liquids at high temperatures which are described in detail elsewhere \cite{Kob95,TWLB02}:
\begin{itemize}
\item 
The already mentioned KA model \cite{Kob95} for binary mixtures of LJ beads in $d=3$ has been investigated 
by means of Langevin MD simulation  \cite{FrenkelSmitBook} imposing a temperature $T=0.8$
for $n=\nA+\nB=6912$ beads per simulation box, a total density $\rho = 1.0$ and 
molar fractions $\ca=\nA/n=0.8$ and $\cb=\nB/n=0.2$ for both types of beads $A$ and $B$.
As in Ref.~\cite{Kob95} we set $\sigAA=1.0\sigma$, $\sigBB=0.88 \sigma$ and $\sigAB=0.8 \sigma$
for the interaction range and $\epsAA=1.0 \epsilon$, $\epsBB=0.5\epsilon$ and $\epsAB=1.5 \epsilon$
for the LJ energy scales. Only data for the usual cutoff $\scut=2.5$ is presented.
\item
Using Monte Carlo (MC) simulations \cite{AllenTildesleyBook,FrenkelSmitBook} we have computed in $d=2$ 
dimensions a specific case of the generalized LJ potential, Eq.~(\ref{eq_LJgeneralized}), where all 
interaction energies are identical, $\epsij= \epsilon$, and the interaction range is set by 
the arithmetic mean $\sigmaij=(D_i+D_j)/2$ of the diameters $D_i$ and $D_j$ of the interacting beads. 
Following Ref.~\cite{TWLB02} the bead diameters are uniformly distributed between $0.8\sigma$ 
and $1.2\sigma$.  For the examples reported here we have used a temperature $T=1.0$, 
$n=10000$ beads per box and a density $\rho \approx 0.72$.
\end{itemize}
We use LJ units throughout this work and Boltzmann's constant $\kB$ is set to unity.
For the indicated parameter choices both systems correspond to isotropic liquids.
The Table summarizes various properties for polydisperse LJ beads for different reduced cutoff 
distances $\scut/\smin$. 
Considering thermodynamic properties per particle (rather than per volume), we have made the
data dimensionless by rescaling with the inverse temperature $\beta$ and the density $\rho$.

\section{Computational results}
\label{sec_res}

\begin{figure}[t]
\centerline{\resizebox{0.95\columnwidth}{!}{\includegraphics*{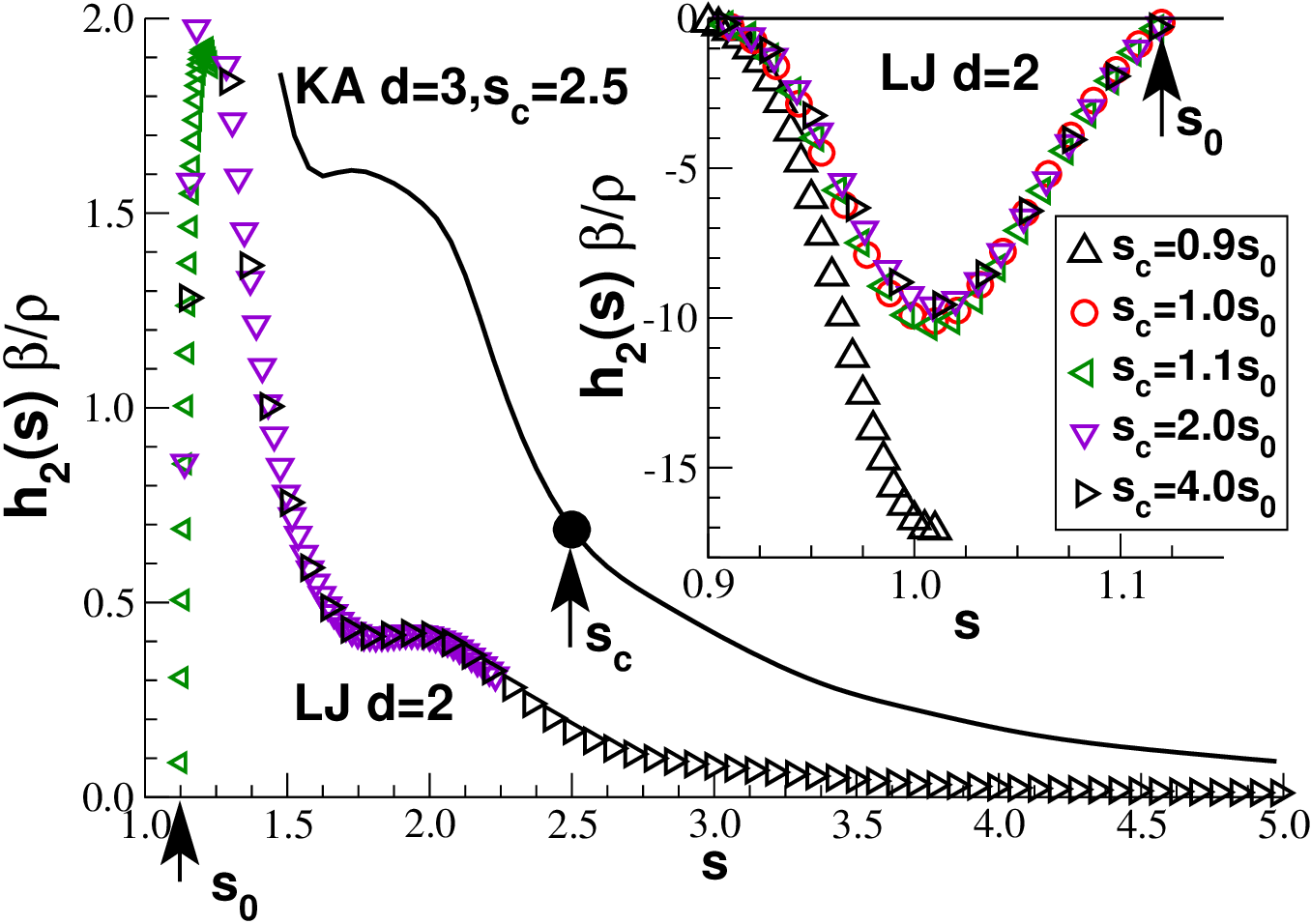}}}
\caption{Weighted radial pair distribution function $\histoB(s) \beta/\rho$ with $s = r/\sigmaij$ 
being the reduced distance between two beads $i$ and $j$.
Main panel: KA mixtures in $d=3$ (bold line) and polydisperse LJ beads in $d=2$ (open symbols) 
for large reduced distances $s > \smin$ where the potential is attractive. 
The filled sphere corresponds to the shear modulus $\Guncorr$ computed using Eq.~(\ref{eq_lame2G}) 
for the KA system not taken into account the impulsive correction.
Inset: Polydisperse LJ beads for $s \le \smin$ where $\histoB(s)$ becomes strongly negative.
\label{fig_histo}
}
\end{figure}

\subsection{Weighted pair distribution function $\histoB(s)$}
\label{res_histo}
The weighted pair distribution function $\histoB(s)$, Eq.~(\ref{eq_histoLame}), is presented in Fig.~\ref{fig_histo}.
Several cutoff distances $\scut$ are given for the polydisperse LJ model, but for clarity only
for distances $s \le \scut$. For the KA model only one cutoff is given,
but this also for $s > \scut$. Note that albeit different $\scut$ for each model correspond strictly speaking
to different state points 
--- as better seen from the energies per bead $e$ or total pressures $P$ indicated in the Table ---
the histograms vary only weakly with $\scut$. Strong differences become only apparent for
very small $\scut$ as shown for $\scut = 0.9 \smin$ in the inset.
One can thus use the histogram obtained for one $\scut$ to anticipate the impulsive
correction for a different cutoff. 
Note that for large distances corresponding to an attractive interaction we have $\histoB(s) > 0$ (main panel). 
Obviously, $\histoB(s)$ vanishes at the minimum of the potential $s=\smin$ and for very large distances $s$.
Since $g(s) \approx 1$ in the latter limit, the histogram $\histoB(s)$ is given 
(up to a known prefactor) by $s^{d+1} \ushiftprime(s)$. 
As one expects the decay is faster for the $d=2$ data than for the KA mixtures in $d=3$,
since the phase volume at the cutoff is larger for the latter systems.
Since all histograms are rather smooth, one may simply set $s = \scut$ for obtaining $\muBcut$
from $\histoB(s)$ instead of properly taking the limit $s \to \scutm$.


\subsection{Compression modulus $K$}
\label{res_compress}
The compression modulus $K$ may be obtained from Eq.~(\ref{eq_lame2K}) or, equivalently, 
using the Rowlinson formula given elsewhere \cite{AllenTildesleyBook,papthermoflex,foot_Kt}. 
All our systems are highly incompressible, i.e. the compression modulus $K$ is large as usual 
in condensed matter systems, and it is thus difficult to demonstrate the small correction predicted by 
Eq.~(\ref{eq_deltaK}). For the KA model we obtain, e.g.  
$\Kcut \beta/\rho \approx (5/4) \times 0.69 \approx - 1.2$
which compared to the uncorrected estimate $\Kuncorr \beta/\rho \approx 21.9$ is not very impressive.

More importantly, it is not easy to obtain an independent and precise $K$-value
for canonical ensembles of mixtures and polydisperse systems using, e.g., the total particle structure factor 
\cite{HansenBook,papthermoflex}. For polydisperse LJ beads we have thus computed $K$ directly from 
the volume fluctuations $\delta V$ in the isobaric ensemble \cite{AllenTildesleyBook}
\begin{equation}
K = \kBT \frac{\la V \ra}{\la \delta^2 V\ra}
\label{eq_Kisobaric}
\end{equation}
where we impose the same (mean) pressure $P$ as for the corresponding canonical ensemble.
As may be seen from the last column indicated in the Table, this yields similar values as the stress-fluctuation formula, 
Eq.~(\ref{eq_lame2K}). Unfortunately, for larger cutoffs our error bars become too large to confirm the correction.
The most striking example, where Eq.~(\ref{eq_deltaK}) can be shown to work, is the case of the 
small cutoff $\scut = 0.9 \smin$: Using Eq.~(\ref{eq_lame2K}) an impossible {\em negative} value 
$\Kuncorr \beta/\rho \approx -19.2$ is obtained.
As may be seen from the inset in Fig.~\ref{fig_histo}, one gets 
$\muBcut \beta/\rho \approx 17.1$ from the weighted 
histogram $\histoB(s)$. Taking the correction Eq.~(\ref{eq_deltaK}) into account this yields 
$K \beta/\rho \approx 15$ which is similar to the value obtained using Eq.~(\ref{eq_Kisobaric}). 


\begin{figure}[t]
\centerline{\resizebox{0.95\columnwidth}{!}{\includegraphics*{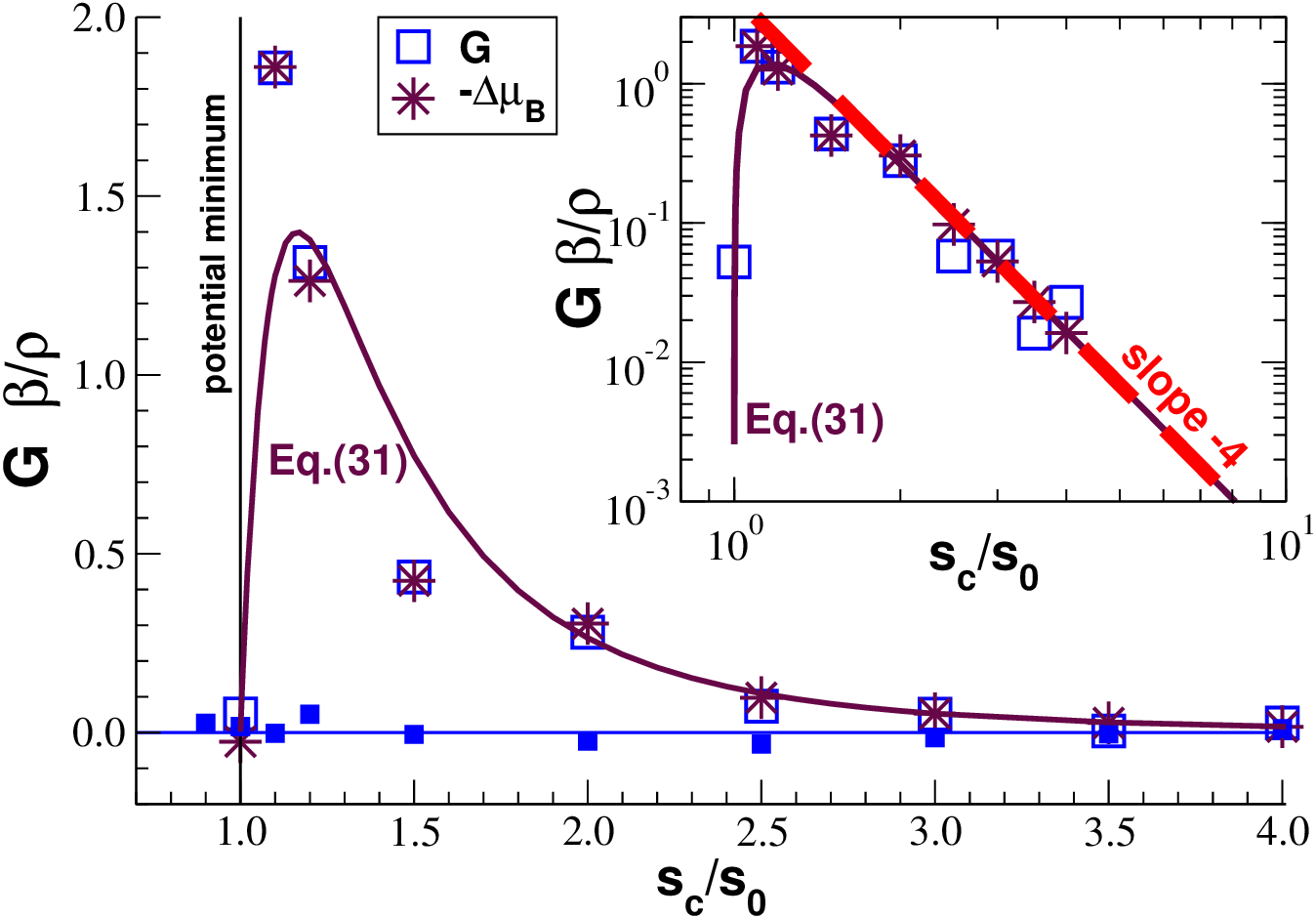}}}
\caption{Shear modulus $G$ and impulsive correction $-\muBcut$ for polydisperse LJ beads 
{\em vs.} the reduced cutoff distance $\scut/\smin$. The uncorrected shear modulus $\Guncorr$
(open symbols) has been obtained using the stress-fluctuation formula, Eq.~(\ref{eq_lame2G}), 
the correction term (stars) from the histogram $\histoB(s)$, Eq.~(\ref{eq_histoLame}). 
The solid lines indicate Eq.~(\ref{eq_muBcut_polyLJ}) where we have set $\gDiDj(\scut) = 1$.
Main panel: Linear representation showing that $G = \Guncorr + \muBcut$ (filled squares)
vanishes as predicted, Eq.~(\ref{eq_deltaG}).
Inset: Double-logarithmic representation of the same data emphasizing the
asymptotic power-law decay for large $\scut$ as indicated by the bold dashed line.
\label{fig_Gscut}
}
\end{figure}

\subsection{Shear modulus $G$}
\label{res_shear}

\paragraph*{Asymptotic limit for large sampling times.}
Since all our systems are liquids, the shear modulus $G$ should of course vanish
--- at least in the thermodynamic limit for a sufficiently long sampling time. 
We have thus a clear reference and for this reason $G$ is highly suitable to test our predictions.
As can be seen from the solid sphere indicated in Fig.~\ref{fig_histo} for the KA mixtures with 
$\scut=2.5$ we obtain $\Guncorr \beta/\rho \approx 0.65$ if the impulsive correction for the Born term 
is {\em not} taken into account.  As also shown by the figure, this deviation equals 
$\histoB(\scut) \beta/\rho \approx 0.69$ as predicted.
The same behavior is seen from Fig.~\ref{fig_Gscut} for polydisperse LJ beads 
for a broad range of cutoff distances $\scut$ where the open squares refer to the uncorrected $\Guncorr$
and the filled squares to $G$ obtained using Eq.~(\ref{eq_deltaG}). 
The solid lines indicated show Eq.~(\ref{eq_muBcut_polyLJ}). Focusing on the scaling for large $\scut$
we have set $\gDiDj(\scut) =1$ in the double-integral which (under this assumption) is close to unity.

\begin{figure}[t]
\centerline{\resizebox{0.95\columnwidth}{!}{\includegraphics*{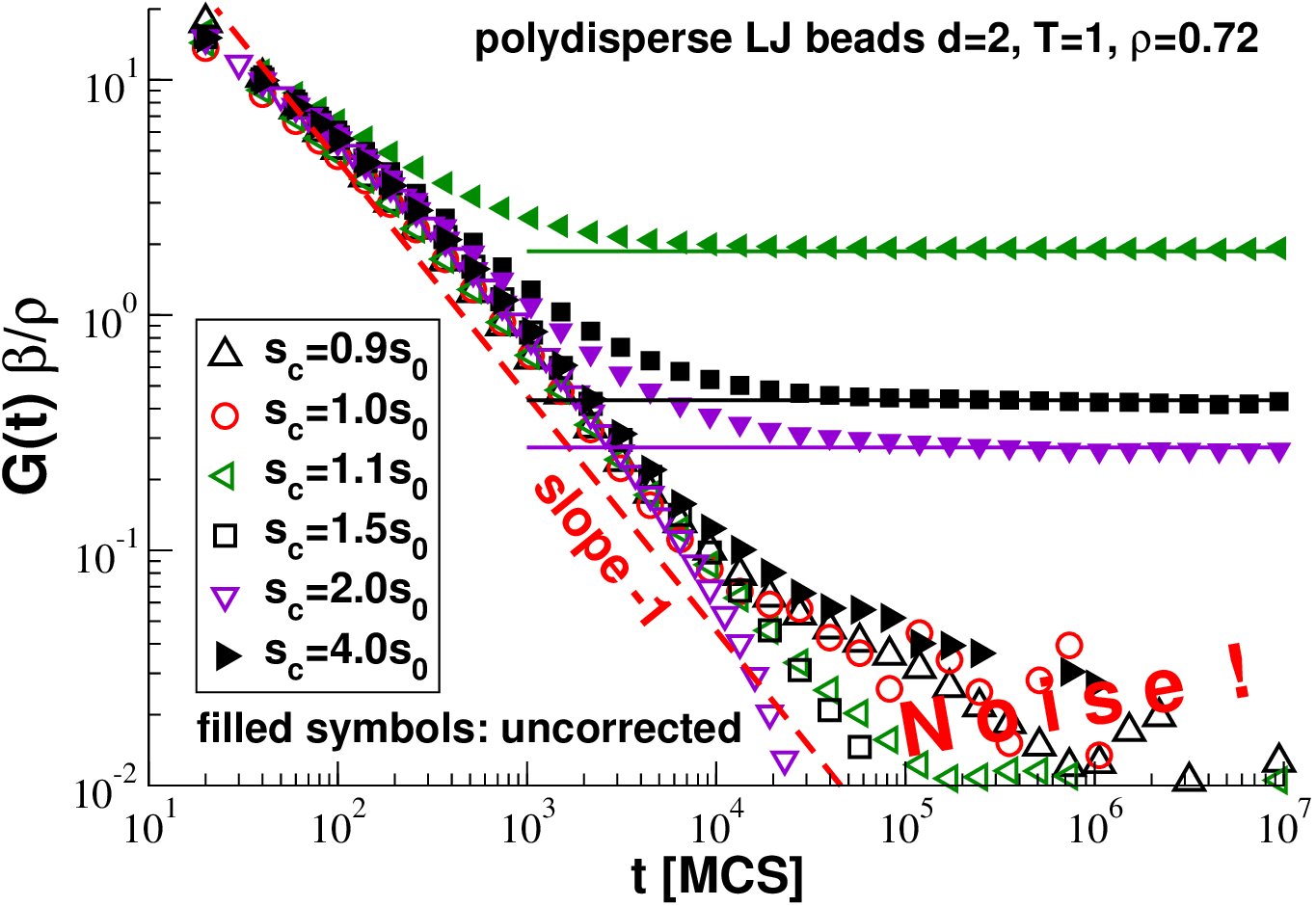}}}
\caption{Shear modulus $G$ for polydisperse LJ beads in $d=2$ for different $\scut$ as a 
function of the sampling time $t$ given in units of MC Steps (MCS) of the local MC jumps used. 
The vertical axis is made dimensionless by means of a factor $\beta/\rho$.
Filled symbols refer to the uncorrected $\Guncorr(t)$.
The horizontal lines indicate $-\muBcut$ obtained from the histograms $\histoB(s)$ for three cutoffs
as indicated in the Table.
The dashed slope characterizes the decay of (the corrected) $G(t)$ with time.
\label{fig_Gt}
}
\end{figure}

\paragraph*{Sampling time dependence.}
Figure \ref{fig_Gt} gives additional information for the shear modulus $G(t)$ plotted as a function 
of the number $t$ of MC steps (MCS) for polydisperse LJ beads. Note that the sampling time is proportional 
to the number of configurations used for the averages. 
To smooth the data we have used gliding averages of length $t$ over the total trajectories of length $10^{7}$ MCS. 
For $\scut=1.0\smin$ and $\scut=4.0\smin$ there was no need to add a correction while for $\scut=0.9\smin$ 
where $\muBcut \beta/\rho \approx 17.1$ the uncorrected data is negative and cannot be represented. 
The filled symbols refer to the uncorrected shear modulus $\Guncorr(t)$ for $\scut = 1.1 \smin$, 
$\scut = 1.5 \smin$ and $\scut = 2.0 \smin$ which are seen to approach for large times the predicted
correction $-\muBcut$ taken from the Table (horizontal lines). 
If corrected, all data sets vanish properly with time.

Interestingly, neither $\muB$ nor $\Pex$ do (essentially) depend on $t$ while the fluctuation 
contribution $-\muF(t)$ approaches (the corrected) $\muB-\Pex$ from below (not shown).
The (corrected) shear modulus $G(t)$ thus decreases monotonously with time. 
As can be seen from Fig.~\ref{fig_Gt}, the (corrected) $G(t)$ decays roughly 
as the power-law slope $-1$ indicated by the dashed line. 
(Note that the noise becomes too large for $G(t)\beta/\rho < 0.1$.)
Exactly the same behavior has been observed for the KA model in $d=3$ (not shown).
Apparently, $G(t)$ decays quite generally inversely 
as the mean-square displacement $\msd(t)$ of the beads in the free-diffusion limit, $\msd(t) \sim t$.
We remind that the same scaling $G(t) \sim 1/\msd(t)$ has also been reported for a bead-spring polymer model 
without impulsive corrections ($\scut = \smin$) \cite{papthermoflex}.

\section{Conclusion}
\label{sec_conc}

\paragraph*{Summary.}
It has been emphasized in this study that an impulsive correction to the Born contributions $\CBabcd$ of 
the elastic moduli must arise if the interaction potential is truncated and 
shifted, Eq.~(\ref{eq_ushift}), with a non-vanishing first derivative at the cutoff.
To test our theoretical predictions we have computed the elastic moduli of isotropic liquids 
in $d=3$ and $d=2$ dimensions. Since for these systems the shear modulus $G$ must vanish by construction, 
this allows a precise numerical verification for different reduced cutoff distances $\scut$. 
It has been shown how the impulsive correction for mixtures and polydisperse systems may be obtained 
from the readily computed weighted histogram $\histoB(s)$ which scales
as $\histoB(s) \sim s^{d+1}\uprime(s)$ for large $s$. 
As one expects, the cutoff effect vanishes if $\scut$ is large, Eq.~(\ref{eq_muBcut_asymp}),
or set to a minimum of the potential. It becomes more important with increasing dimension. 

\paragraph*{Comment.}
Incidentally, it should be noted that the stress-fluctuation formula $G = \muB + \muF -\Pex$
and various other relations used in this work for {\em liquid} systems were originally derived 
for {\em solids} assuming well-defined reference positions and displacement fields 
\cite{Hoover69,Ray84,Ray88,Lutsko89,Barrat06}. It can be shown, however, that
these assumptions can be relaxed and especially Eq.~(\ref{eq_lame2G}) holds quite generally
for isotropic systems. 
The aim of the present paper was to show numerically that the stress-fluctuation formalism yields 
the right value ($G=0$) {\em once} the impulsive correction has been taken into account.

\paragraph*{Outlook.}
We are currently using the approach presented here to characterize as a function of temperature,
imposed pressure and sampling time the glass transition of the two models presented here.
The generalization of our results 
to 
\begin{itemize}
\item 
other elastic moduli in anisotropic systems using the more general impulsive correction Eq.~(\ref{eq_CBabd_correct}),
\item
observables related to even higher derivatives of the potential, Eq.~(\ref{eq_hn_histo}), and
\item
arbitrary interaction potentials not necessarily scaling simply with $s = r/\sigmal$ 
and not necessarily being pair interactions using the generalization of the Born term
derived by Ray \cite{Ray88} 
\end{itemize}
is straightforward and will be considered in the future.
%

\begin{acknowledgments}
H.X. thanks the CNRS and the IRTG Soft Matter for supporting her sabbathical stay in Strasbourg
and P.P. the  R\'egion Alsace and the IRTG Soft Matter for financial support.
We are indebted to A. Blumen (Freiburg) and H.~Meyer, O.~Benzerara and J. Farago (all ICS, Strasbourg) 
for helpful discussions.
\end{acknowledgments}



\begin{thebibliography}{31}
\bibitem[{foo({\natexlab{a}})}]{foot_CKabcd}
\bibinfo{note}{Apart the Born term $\CBabcd$ and the stress-fluctuation term
  $\CFabcd$ there is a kinetic contribution $\CKabcd$ and in pre-stressed
  systems (as in the systems considered numerically by us) an explicit
  contribution from the applied stress to the experimentally relevant elastic
  moduli associated to an infinitesimal incremental strain from the reference
  state \cite{FrenkelSmitBook,Wallace70,SBM11}.}
\bibitem[{foo({\natexlab{b}})}]{foot_kinelast}
\bibinfo{note}{The trivial kinetic energy contributions to the elastic moduli
  are removed as far as possible from the presentation in view of the fact that
  Monte Carlo results are considered here.}
\bibitem[{foo({\natexlab{c}})}]{foot_Kt}
\bibinfo{note}{If plotted as a function of the number of configuration sampled,
  the compression modulus for both models is seen to decrease first with
  sampling time $t$ before leveling off at a finite value. Similar behavior has
  been observed for polymeric systems \cite{SBM11,papthermoflex}.}
\end{thebibliography}


\end{document}